# A Privacy Preserving Improvement for SRTA in Telecare Systems

**Seyed Salman Sajjadi Ghaemmaghami[1], Mahtab Mirmohseni[2], Afrooz Haghbin[1]**

**Abstract**

Radio Frequency Identification (RFID) is a modern communication technology, which provides authentication and identification through a nonphysical contact. Recently, the use of this technology is almost developed in healthcare environments. Although RFID technology can prepare sagacity in systems, privacy and security issues ought to be considered before. Recently, in 2015, Li et al. proposed SRTA, a hash-based RFID authentication protocol in medication verification for healthcare. In this paper, we study this protocol and show that SRTA protocol is vulnerable to traceability, impersonation and DoS attacks. So it does not provide the privacy and security of RFID end-users. Therefore, we propose an improved secure and efficient RFID authentication protocol to enhance the performance of Li et al.'s method. Our analyze show that the existing weaknesses of SRTA's protocol are eliminated in our proposed protocol.

## 1. Introduction

Radio Frequency Identification (RFID) technology has outlined a novel future for our world. Aviation, building management, financial services, livestock and animal tracking, marina, passenger transport, supply chain, rail way and health-care are some examples of RFID usages which describe the variety of its application in our life [1-4]. Nowadays, the increased utilization of RFID systems in healthcare has been grown substantially, for instant patient tracking, wait-time monitoring, medication authentication and control asset management, document and file tracking, laundry and waste management can be classified as its applications in this field [5-7]. As shown in Fig. 1, RFID systems consist of three main parts: tag, reader and back-end server. The tag is placed inside the products or the proposed items, for authentication and identification in contact with the readers. Tags are categorized in one of the three classes: active, passive and semi-active. A passive tag does not have any battery, so it cannot start a new connection unless locates in the electromagnetic field of the reader, to gain enough power for transmitting its messages. An active tag normally operates at 433MHz Ultra High Frequency (UHF) and has an inner battery which lets it to start a new conversation with the reader whenever it wants; Of course these properties increase the cost and the volume of this type of tags which constrain its usage in military applications, at microwave and ultra-wide band frequency ranges [8]. A semi-active tag has a battery, which only uses it to perform internal operations; rely on the reader's signal to power their antenna and modulator [9]. The back-end server connects to the readers through the secure or unsecure channels and stores all the identification information of the readers and the tags in its database for further processing.

"98000 people annually die due to medication related mistakes in the United States," reported by the Institute of Medicine (IOM) [10] which is the result of three main facts: similarity in the name of medicine, packing and types of labels [11-13]. Nowadays, in

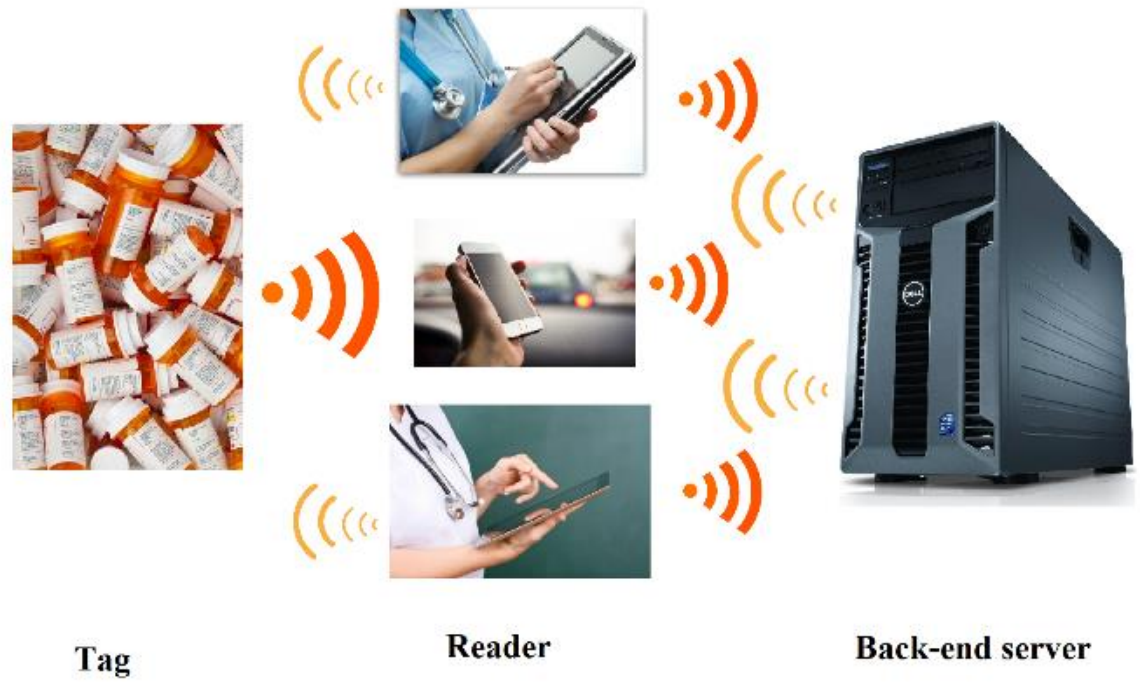

Figure 1. RFID system

Salman.ghaemmaghami@srbiau.ac.ir
mirmohseni@sharif.edu
haghbin@srbiau.ac.ir

1. Department of Computer Engineering, Science and Research branch, Islamic Azad University, Tehran, Iran

2. Department of Electrical Engineering, Sharif University of Technology, Tehran, Iran

order to establish confidentiality and privacy, and solve the problems of existing methods, new protocols have been proposed [13-16]; According to the state of the IOM, a number of those are specifically considered for Telecare Medicine Information System (TMIS) [5, 17, 18]. It is undeniable that an efficient RFID security scheme can increase the security and privacy of RFID end-users significantly [19].

In 2011, Chen et al. [20] proposed a tamper resistant prescription RFID access control protocol for different certified readers where both authentication and access right authorization mechanisms were and it was claimed to guarantee patient's right. In the same year, a new hash-based RFID mutual authentication protocol was proposed by Cho et al. [21]; they believe that their protocol makes it difficult for an attacker to launch an effective brute-force attack against RFID users. But Kim et al. [22] showed that Cho et al.'s protocol is weak against desynchronization attack and proposed a hash-based mutual authentication protocol to solve the security problems in Cho et al.'s protocol and privacy problems in previous RFID authentication protocols. In 2012, Yu et al. proposed a grouping proof protocol [23] for low cost RFID tags and showed that not only the number of logic gates in their protocol was reduced but also it requires fewer computational power and operation costs versus the last proposed protocol. In the same year, Wu et al. [24] showed that Yu et al.'s protocol was still vulnerable to impersonation attacks and proposed a lightweight binding proof protocol to overcome their weaknesses. Srivastava et al. [6] proposed a protocol in 2015 to strengthen the security level of common protocol, using hash algorithm and synchronized secret value shared between the tag and the back-end server; which was believed to be safe against various active and passive attacks. However, Li et al. [7] showed in SRTA (Secure RFID Tag Authentication) protocol that Srivastava et al.'s tag authentication protocol has security problem which let an adversary use the lost reader to connect to the medical back-end server. Moreover, they believe that Srivastava et al.'s protocol fails to provide mutual authentication between the reader and the back-end server, so they have proposed a secure and efficient RFID tag authentication protocol to overcome the mentioned weaknesses.

In this paper, we analyze the SRTA protocol [7] and show that there are still weaknesses with their protocol. Using timestamp in the structure of their protocols was the novelties of Srivastava et al. and Li et al. which prevents data forgery and replay attacks. However, we show that declaring timestamps explicitly through the protocol in one hand and inaccuracy in producing the messages on the other hand, lead to the tag impersonation and reader impersonation attacks. Moreover, expressing the reader and tag's identification values through the authentication phases and lack of appropriate updating procedure put the privacy of their protocol at risk. In order to investigate the privacy of this protocol, we use Ouafi and Phan privacy model [25] and by consuming the mentioned vulnerabilities, we present the tag and reader traceability attacks on SRTA protocol [7]. Besides, it should be known that low cost of RFID's tag results in computation and complexity restrictions in the tag side, but this restriction is not so serious in the back-end server due to the presence of powerful processors [19]. Therefore, we propose an improved version of SRTA protocol [7] that prevents the mentioned attacks and decreases the computation cost in the tag side.

The rest of the paper is organized as follows: the privacy model of Ouafi and Phan is described in Section 2. SRTA protocol is reviewed in Section 3. In Section 4, SRTA protocol is analyzed and its weaknesses are discussed. An improved version of Li et al.'s protocol is proposed in Section 5 and analyzes of our improved version are discussed in Section 6. Finally, the paper is concluded in Section 7.

## 2. Privacy model of Ouafi and Phan

Providing a confidential communication for RFID users is one of the main goals of each RFID communications scheme. As a result, studying privacy of the proposed authentication protocols always is more prominent for researchers [26-28]. In order to evaluate the privacy of RFID protocols, different models have been proposed, and one of the appropriate and well-known model is Ouafi and Phan privacy model [25], which is described in this section. It is an Untraceable Privacy (UPriv) model which can briefly mentioned as follows:

The reader $R$ and the tag $T$ are the components of the model and the communications between all protocol parties are managed by an adversary $\mathcal{A}$, based on the protocol definition. The following queries can be run by an adversary $\mathcal{A}$:

■ **Execute $(R, T, i)$ query**: This query is categorized as passive attack and let the attacker $\mathcal{A}$ eavesdrop the transmitted messages between the reader $R$ and the tag $T$ in the $i$th session of the protocol.

■ **Send $(U, V, i, m)$ query**: An active attack is modeled with this query by sending the message $m$ from the $U \in$ tag $T$ (reader $R$) to the $V \in$ reader $R$ (tag $T$) in the $i$th session of protocol. Besides, the adversary $\mathcal{A}$ can alter or block the exchanged messages.

■ **Corrupt $(T, k)$ query**: The attacker $\mathcal{A}$ is able to obtain $K'$, the secret value of the tag $T$ and set it to $K$.

■ **Test $(T_0, T_1, i)$ query**: This query allows to express the indistinguishability based concept of UPriv. After sending a Test $(T_0, T_1, i)$ query to an entity in the $i$th session, depending on a randomly chosen bit $b \in \{0,1\}$ generated by challenger, $T_b \epsilon \{T_0, T_1\}$ is delivered to the attacker. Adversary $\mathcal{A}$ will succeed, if it can truly guess the bit $b$.

**Untraceable Privacy (UPriv)**: In this definition a game $G$ between the attacker $\mathcal{A}$ and a collected instances of reader and tag is taking place. An adversary $\mathcal{A}$ runs the game $G$ which has the following phases:

∵ **Learning phase:** In this phases, an adversary $\mathcal{A}$ is permitted to send each of Execute, Send and Corrupt queries.

∵ **Challenge phase:** An adversary $\mathcal{A}$ is given a tag $T_b \epsilon \{T_0, T_1\}$ and sends any of Execute, Send and Corrupt queries to $T_b$.

∵ **Guess phase:** Finally, the adversary $\mathcal{A}$ terminates the game $G$ and outputs a bit $b_0$ as a guess of the value

| **Back-end Server** $(ID_k, V_k, W_k, x_j, x_{j-1}, s_j, s_{j-1}, RID_k)$ | **Reader** $(x_j, RID_k)$ | **Tag** $(s_j, ID_k)$ |
|---|---|---|
| | 1 $input\ RID_k\ and\ RPW_k$<br>1.1 $V_k = h(x_j \parallel RID_k)$<br>1.2 $W_k = h(x_j \parallel RID_k) \oplus RID_k \oplus RPW_k$<br>1.3 $V'_k = W_k \oplus RID_k \oplus RPW_k$<br>$if\ V_k = V'_k$<br>1.4 Generates $R_r$<br>1.5 $A = V'_k \oplus R_r$<br>1.6 $B = h(V'_k \oplus T_1 \oplus R_r)$<br>1.7 $\xrightarrow{A,B,RID_k,T_1}$ | |
| | | 2.1 Generates $R_t$ Randomly<br>2.2 $C = h(s_j \parallel ID_k) \oplus R_t$<br>2.3 $D = h(h(s_j \parallel ID_k) \oplus T_2 \oplus R_t)$<br>2.4 $\xleftarrow{C,D,ID_k,T_2}$ |
| | 3.1 $if\ T_2 - T_1 > \Delta T$<br>$Reveal\ the\ Protocol$<br>$else$<br>3.2 $\xleftarrow{A,B,RID_k,T_1,C,D,ID_k,T_2}$ | |
| 4.1 $if\ T_3 - T_2 > \Delta T$<br>$Reveal\ the\ Protocol$<br>$else\ for\ each\ tuple\ (x_j, x_{j-1})$<br>4.2 $computes\ R_r^*, B^*$<br>$for\ each\ tuple\ (s_j, s_{j-1})$<br>4.3 $computes\ R_t^*, D^*$<br>4.4 $E = h(x_j \parallel RID_k \parallel T_1 \parallel R_r^* \parallel h(x_j \oplus R_r^*))$<br>4.5 $F = Data \oplus \mathrm{h}(x_j \oplus R_r^*)$<br>4.6 $G = h(s_j \parallel ID_k \parallel T_2 \parallel R_t^* \parallel h(s_j \oplus R_t^*))$<br>4.7 $\xrightarrow{E,F,G}$ | | |
| 4.8 After successful authentication updates<br>$x_{j-1} \leftarrow x_j;\ x_j \leftarrow h(x_j \oplus R_r)$<br>$s_{j-1} \leftarrow s_j;\ s_j \leftarrow h(s_j \oplus R_t)$ | 5.1 $Compute$<br>$E^* = h(x_j \parallel RID_k \parallel T_1 \parallel R_r \parallel h(x_j \oplus R_r))$<br>5.2 $Check\ E^* \stackrel{?}{=} E$<br>5.3 Updating $x_j \leftarrow h(x_j \oplus R_r)$<br>5.4 $Data = F \oplus \mathrm{h}(x_j \oplus R_r)$<br>5.5 $\xrightarrow{G}$ | |
| | | 6.1 $Compute$<br>$G^* = h(s_j \parallel ID_k \parallel T_2 \parallel R_t \parallel h(s_j \oplus R_t))$<br>6.2 $Verify\ G^* \stackrel{?}{=} G$<br>6.3 $After\ successful\ authentication$<br>$s_j \leftarrow h(s_j \oplus R_t)$ |

Fig. 2 The SRTA protocol [7].

of $b$.

The attacker $\mathcal{A}$ is succeeded during playing the game $G$, if it recognizes correctly whether received $T_0$ or $T_1$. The traceability level of the protocol is denoted by $Adv_A^{UPiv}(\kappa)$, where $k$ is the security parameter:

$$Adv_A^{UPiv}(k) = |\text{pr}(\mathcal{A}\ wins) - \text{pr}(random\ coin\ flip)| = |\text{pr}(b' = b) - \tfrac{1}{2}| \quad (1)$$

where $0 \leq Adv_A^{UPiv}(k) \leq \frac{1}{2}$. If $Adv_A^{UPiv}(k) < \varepsilon(k)$, the protocol is traceable with negligible probability.

## 3. SRTA Protocol

In [7], Li et al. proposed a secure RFID tag authentication protocol in TMIS. The connection between the reader and the back-end server and the connection between the tag and the reader is insecure. Their protocol is a hash based one, which uses timestamps in the structure of its messages to prevent attacks. Their protocol is depicted in Fig. 2 and notations that are used in this protocol are listed below:

$ID_k$: The identifier of the $k$th tag.
$RID_k$: The identifier of the $k$th reader.
$RPW_k$: The password of the $k$th reader.
$RNG$: The Random Number Generator.
$T$: The timestamp.
$R_r$: The random number generated by reader.
$R_s$: The random number generated by tag.
$s_j$: The secret value used in the current $j$th session and it is mutual shared between back-end server and tag.
$s_{j-1}$: The secret value used in the previous session. Initially, the value is set to null.
$x_j$: The secret value used in the current $j$th session and it is mutual shared between back-end server and reader.
$x_{j-1}$: The secret value used in the previous session. Initially, the value is set to null.
$h(.)$: A one-way hash function.
$\Delta T$: The expected legitimate time interval for transmission delay.
$\|$: Concatenation operation.
$A \oplus B$: Message A is XORed with message B.

## 4. Analyzes of SRTA Protocol

### a. Tag Impersonation

Li et al. try to increase the security in authentication procedure by using timestamps, which means that the reader and the back-end server will not continue the authentication phase, unless the inequalities $\{T_2 - T_1 < \Delta T, T_3 - T_2 < \Delta T\}$ occurred. So by knowing the values of $\Delta T$ and $T_1$, $T_2$ and $T_3$, the attacker tries to impersonate a legitimate tag to receive responses from the reader. It is shown that an attacker can perform this attack on Li et al.'s protocol [6]. This attack can be performed as follows,

***Learning phase:*** In the $i$th round, the attacker eavesdrops four successful steps of the protocol and obtains $\{RID_k, A, B, T_1, ID_k, C, D, T_2\}$ and by changing $T_2$ into $T'_2$, in which $T'_2 - T_1 > \Delta T$, he/she leaves the protocol unfinished. So the secret values of the reader and the tag are not updated.

***Attack phase:*** In the $(i + 1)$th round, the attacker starts a new session with the reader and acts as follows,

a) The attacker $\mathcal{A}$ receives $\{RID_k, A, B, T_1^{(i+1)}\}$ from the reader. By knowing the value of $T_1^{(i+1)}$ in this session and $\Delta T$ from the learning phase, he/she generates the $T_2^{(i+1)}$ and responses with $\{ID_k, \beta, \gamma, T_2^{(i+1)}\}$ that $\beta$ and $\gamma$ are generated as follows,

$$\beta = h(s_j \parallel ID_k) \oplus R_t \quad (2)$$

$$\gamma = h(h(s_j \parallel ID_k) \oplus T_2^{(i+1)} \oplus R_t) \quad (3)$$

b) After confirming the value of $T_2^{(i+1)}$ by calculating $\Delta T$ in the reader side, $\{RID_k, A, B, T_1^{(i+1)}, ID_k, \beta, \gamma, T_2^{(i+1)}\}$ will be sent to the back-end server by the reader.

c) By receiving the response messages from the reader, the back-end server checks for the inequality $(T_3^{(i+1)} - T_2^{(i+1)}) < \Delta T$ which will be accepted by choosing a correct value for $T_2^{(i+1)}$ via the attacker. As the above inequality holds, the back-end server acts as follows:

   1. Computes $R_r^* = A \oplus h(x_j \parallel RID_k)$.
   2. Computes $B^* = h(h(x_j \parallel RID_k) \oplus T_1^{(i+1)} \oplus R_r^*)$ and checks if $B^* \stackrel{?}{=} B$. So the back-end server successfully authenticates the reader.
   3. Computes $R_t^* = \beta \oplus h(s_j \parallel ID_k)$.
   4. Computes $\gamma^* = h(h(s_j \parallel ID_k) \oplus T_2^{(i+1)} \oplus R_t^*)$ and checks if $\gamma^* \stackrel{?}{=} \gamma$. As the secret value of the tag has not been updated, the above equality is confirmed.

Therefore, the back-end server authenticates the attacker as a legitimate tag.

### b. DoS Attack

It can be shown that Li et al.'s protocol is not safe against DoS attack. To perform this attack, in the $i$th session of the protocol, after running four steps, when the back-end server wants to send messages to the reader, the attacker $\mathcal{A}$ intercepts the transmitted messages and stops the protocol. As a result, the back-end server updates $s_j^{(i)}$ and $s_{j-1}^{(i)}$ with $h(s_j \oplus R_t)$ and $s_j$, respectively, but the tag dose not update its secret values. Now, the attacker $\mathcal{A}$ performs the tag impersonation attack, presented in Section 4.a, in $(i+1)$th session of the protocol. After this attack, the back-end server updates $s_j^{(i+1)}$ and $s_{j-1}^{(i+1)}$ with $h(s_j^{(i)} \oplus R_t)$ and $s_j^{(i)}$, respectively, but the tag dose not update its secret values. Consequently the tag and the back-end server are desynchronized in the next session and the back-end server cannot authenticate the tag.

In addition, the DoS attack can be performed by running two consecutive tag impersonation attacks, described in subsection 4.a.

## c. Reader Impersonation

In this subsection, it is shown that an attacker can impersonate a legitimate reader in Li et al.'s protocol [7]. This attack can be performed as follows:

***Learning phase:*** In the $i$th round, the attacker eavesdrops two successful steps of the protocol and obtains $\{RID_k, A, B, T_1\}$, intercepts the transmitted messages to the tag and then stops the protocol. So the secret values are not updated in this session. The attacker calculates $\alpha$ as follows:

$$\alpha = V_k' \oplus R_r \quad (4)$$

***Attack phase:*** In the $(i+1)$th round, an adversary $\mathcal{A}$ starts a new session with the tag $T_0$ and acts as follows:

a) In this phase, the attacker starts a session with a tag by sending $RID_k$ and $\alpha$, stored from the last an unfinished session. $T_1^{(i+1)}$ generated by the attacker $\mathcal{A}$ which shows the current timestamp and $\lambda$ which is calculated as

$$\lambda = h(V_k' \oplus T_1^{(i+1)} \oplus R_r) \quad (5)$$

b) Then, the target tag responds $\{ID_k, C, D, T_2^{(i+1)}\}$ to the attacker.

c) The attacker $\mathcal{A}$ sends $\{RID_k, \alpha, \lambda, T_1^{(i+1)}, ID_k, C, D, T_2^{(i+1)}\}$ to the back-end server .

d) The back-end server checks if $(T_3^{(i+1)} - T_2^{(i+1)}) < \Delta T$. As shown in Fig. 2, this inequality is verified because of generation of $T_2^{(i+1)}$ and $T_3^{(i+1)}$ by a legal tag and back-end server.

e) By performing the above steps, the back-end server computes $R_r^* = \alpha \oplus h(x_j \parallel RID_k)$.

f) The back-end server calculates $B^* = h(h(x_j \parallel RID_k) \oplus T_1^{(i+1)} \oplus R_r^*)$ and checks whether $B^* \stackrel{?}{=} \lambda$ where

$$\begin{aligned} B^* &= h\big(h(x_j \parallel RID_k) \oplus T_1^{(i+1)} \oplus R_r^*\big) \\ &= h\big(V_k' \oplus T_1^{(i+1)} \oplus R_r^*\big) \\ &= \lambda \end{aligned} \quad (6)$$

As a result, the back-end server authenticates the spoofed reader as a legitimate one.

g) Now, the back-end server starts to authenticate the tag by calculating $C^*$ and $D^*$ and comparing them with the received $C$ and $D$. As the tag is legitimate, so the back end server authenticates it and computes $E$, $F$ and $G$ as follows and sends them to the attacker:

$$E = h(x_j \parallel RID_k \parallel T_1^{(i+1)} \parallel R_r \parallel \mathrm{h}(x_j \oplus R_r)) \quad (7)$$

$$F = Data \oplus h(x_j \oplus R_r) \quad (8)$$

$$G = h(s_j \parallel ID_k \parallel T_2^{(i+1)} \parallel R_t \parallel \mathrm{h}(s_j \oplus R_t)) \quad (9)$$

h) The attacker $\mathcal{A}$ sends $G$ to the tag.

Consequently, the attacker effectively impersonate the reader.

## d. Tag traceability

In this subsection, it is shown that SRTA protocol [7] is vulnerable against traceability attack. According to SRTA protocol [7], it can be seen that the tag's identification number $ID_k$ is fixed in all rounds. Using this issue, an attacker can trace the target tag. This attack is performed as follows:

***Learning phase:*** In round $(i)$, the attacker $\mathcal{A}$ eavesdrops all transmitted messages between the tag $T_0$ and the reader $R$ by sending an Execute query $(R, T_0, i)$ and obtaining $\{RID_k, A, B, T_1, ID_k, C, D, T_2, E, F, G\}$.

***Challenge phase:*** The adversary $\mathcal{A}$ selects two fresh tags $T_0$ and $T_1$ for test, and sends a Test query$(T_0, T_1, i+1)$. According to the randomly chosen bit $b \in \{0,1\}$, the adversary is given a tag $T_b \in \{T_0, T_1\}$. Afterwards, the adversary $\mathcal{A}$ calculates $B^{\#}$ as $h(A \oplus T_1')$ and sends an Execute query$(R, T_b, i+1)$ by sending $RID_k$, $A$, $B^{\#}$, $T_1'$ to the tag ,which $T_1'$ is the current timestamp, and obtains $C'$, $D'$, $T'_2$ and $ID_k$.

***Guess phase:*** The adversary $\mathcal{A}$ stops the game $G$, and outputs a bit $b' \in \{0,1\}$ as a guess of bit $b$ as follows.

$$b' = \begin{cases} 0 & if\ C = C' \\ 1 & otherwise \end{cases} \quad (10)$$

As a result, it can be written:

$$Adv_A^{upriv}(k) = \left|pr(b' = b) - \tfrac{1}{2}\right| = \left|1 - \tfrac{1}{2}\right| = \tfrac{1}{2} \gg \varepsilon \quad (11)$$

***Proof:*** According to the structure of SRTA protocol [7], since the tag $T_0$ does not ever update its identification number and uses the same $ID_k$ in both learning and challenge phases, the attacker can trace the target tag. Moreover, as $ID_k$ is fixed in all sessions, the attacker $\mathcal{A}$ is able to trace the tag $T_0$, whenever he/she wants.

### e. Reader traceability Attack on SRTA Protocol

Li et al. [7] distinguished that Srivastava et al.'s protocol [6] suffers from reader stolen/lost attack, so it fails in providing the privacy of tag during the authentication phases. To resist these attacks, Li et al. [7] use a secret value, identifier and a password for reader in their protocol. In this subsection, it is shown that in Li et al.'s protocol, an attacker can perform traceability attack and traces the location of a specific reader. As shown in Fig. 1, the adversary $\mathcal{A}$ can trace the reader $R_0$ as follows:

***Learning phase:*** In round $(i)$, the attacker $\mathcal{A}$ eavesdrops all transmitted messages between the tag $T_0$ and the reader $R_0$ by sending an Execute query $(R_0, T_0, i)$, obtaining $\{RID_k, A, B, T_1, ID_k, C, D, T_2, E, F, G\}$, then he/she stores $RID_k$ as $\zeta$.

***Challenge phase:*** The adversary $\mathcal{A}$ eavesdrops every sessions between readers and tags and stores all the obtained $RID_k$ with the name of $Z^i$, where $i\epsilon\{1,2,\ldots, number\ of\ Readers\}$. Afterwards, the adversary $\mathcal{A}$ selects two fresh readers $R_0$ and $R_1$ for test, and sends a Test query$(R_0, R_1, i+1)$. According to the randomly chosen bit $b\ \epsilon\ \{0,1\}$, the adversary is given a reader $R_b\ \epsilon\ \{R_0, R_1\}$. Now the attacker sends an Execute query $(R_0, T_0, i+1)$ and stores $Z^0$ and $Z^1$.

***Guess phase:*** The adversary $\mathcal{A}$ stops the game $G$, and outputs a bit $b'\ \epsilon\ \{0, 1\}$ as a guess of bit $b$ as follows:

$$b' = \begin{cases} 0 & if\ \zeta = Z^0 \\ 1 & otherwise \end{cases} \tag{12}$$

As a result, it can be written:

$$Adv_A^{upriv}(k) =$$

$$= \left|pr(b' = b) - \tfrac{1}{2}\right| = \left|1 - \tfrac{1}{2}\right| = \tfrac{1}{2}\varepsilon \tag{13}$$

***Proof:*** According to the structure of Li et al.'s protocol, the reader $R_0$ will not update its identification number and uses the same $RID_k$ in both Learning and Challenge phases, therefore the attacker can trace the target reader. Furthermore, as $RID_k$ is fixed in all rounds, an adversary $\mathcal{A}$ is able to trace the reader $R_0$ in every arbitrary session.

## 5. Improvements on SRTA Protocol

Li et al. [7] try to improve the Srivastava et al.'s authentication protocol [6] by adding the secret value of the reader $x_j$, the Kth reader identifier and password which are named, respectively, by $RID_k$ and $RPW_k$. However, SRTA protocol [7] is vulnerable to attacks declared in Section 4. In this Section, a strengthened versions of SRTA protocol [7] is proposed to overcome its weaknesses. Also, the security and privacy analysis of our proposed protocol is provided.

### 5.1 Improved Version of SRTA protocol

As reported in Section 4, there are several main drawbacks in the structure of the Li et al.'s protocol [7], which make it vulnerable to traceability attacks. Li et al. [7] try to increase the efficiency of the Srivastava et al.'s protocol [6] by expressing the tag's identifier $ID_k$ and $RID_k$ through the protocol, explicitly. Although SRTA protocol [7] decreases the waiting time for accessing the true readers and ensuring a high rate of efficiency in the tag authentication procedure, but it brings a drawback which ables the attacker to know the tag and reader's identification value. This leads to trace them in every execution of the protocol. In addition, the processors in the tags are limited and all computations cannot be performed in the tag side. On the other hand, there is little limitation for the computation cost in the back-end server side [19]. Therefore, we propose to omit sending $ID_k$ through the protocol. Besides, there is not any inconsistency between the increased time for finding a correct $ID_k$ and $RID_k$ with the timestamp $T_3$. In other words, in SRTA protocol [7], the back-end server first investigates the correctness of an inequality $(T_3 - T_2 < \Delta T)$, then explores for the true identification number of the reader and the tag. Further, we omit sending $RID_k$ through our protocol. One of the other drawbacks of SRTA protocol [7] is announcing the value of timestamps $T_1$, $T_2$ and $T_3$, through the protocol. After one run of the protocol acceptably, an adversary $\mathcal{A}$ knows the value of $T_1$, $T_2$ and $T_3$, so he/she can calculate the allowable $\Delta T$ and applying the tag impersonation and reader impersonation attack which are discussed in Section 4. In order to improve Li et al.'s protocol [7], we change the message $B$ to:

$$B = h(\mathrm{R}(V_k') \parallel \mathrm{L}(R_r) \oplus T_1) \tag{14}$$

where $\mathrm{R}(V_k')$ means the right side of $V_k'$ and $\mathrm{L}(R_r)$ refer to the left side of $R_r$. By omitting $T_1$, we send $\{RID_k, A, B\}$ to the tag in the second step of the protocol. In the third step of the protocol, we change the message $D$ to:

$$D = h(R_t \oplus T_2) \tag{15}$$

Not only by omitting the first hash function of the message $D$, the computation cost in the tag side decreases, but also the back-end server can verify the value of $R_t$ using the transmitted message $D$. Moreover, in our proposed protocol the attacker will not be able to guess the correct message.

On the other hand, updating the tag's identifier $ID_k$

**Back-end Server** $(ID_k^{old}, ID_k^{new}, V_k, W_k, x_j, x_{j-1}, s_j, s_{j-1}, RID_k^{old}, RID_k^{new})$ 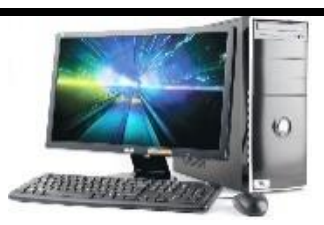

**Reader** $(x_j, RID_k)$ 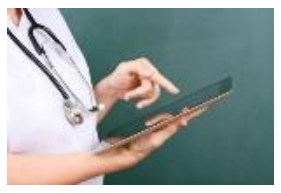

**Tag** $(s_j, ID_k)$ 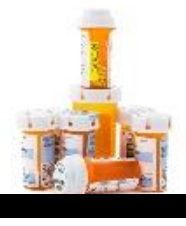

**Reader:**

1 $input\ RID_k\ and\ RPW_k$

1.1 $V_k = h(x_j \parallel RID_k)$

1.2 $W_k = h(x_j \parallel RID_k) \oplus RID_k \oplus RPW_k$

1.3 $V_k' = W_k \oplus RID_k \oplus RPW_k$

$if\ V_k = V_k'$

1.4 $Generates\ R_r$

1.5 $A = V_k' \oplus R_r$

1.6 $B = h(\mathrm{R}(V_k') \parallel \mathrm{L}(R_r) \oplus T_1)$

1.7 $\xrightarrow{A,B}$

**Tag:**

2.1 $Generates\ R_t\ Randomly$

2.2 $C = h(s_j \parallel ID_k) \oplus R_t$

2.3 $D = h(R_t \oplus T_2)$

2.4 $\xleftarrow{C,D,T_2}$

**Reader:**

3.1 $if\ T_2 - T_1 > \Delta T$

$Reveal\ the\ Protocol$

$else$

3.2 $\xleftarrow{A,B,T_1,C,D,T_2}$

**Back-end Server:**

4.1 $if\ T_3 - T_2 > \Delta T$

$Reveal\ the\ Protocol$

$else$

$for\ each\ (x_j, RID_k^{old})\ and\ (x_{j-1}, RID_k^{new})$

4.2 $computes\ V_k^*$

$computes\ R_r^*$

$computes\ B^*$

4.3 $if\ B^* = B$

$Reader\ is\ authenticated$

$else\ reveal\ the\ protocol$

$for\ each$

$(s_j, ID_k^{old})\ and\ (s_{j-1}, ID_k^{new})$

4.4 $computes\ R_t^*, D^*$

4.5 $if\ D^* = D$

$Tag\ is\ authenticated$

$else\ reveal\ the\ protocol$

4.6 $E = h(x_j \parallel RID_k \parallel T_1 \parallel R_r^* \parallel h(x_j \oplus R_r^*))$

4.7 $F = Data \oplus \mathrm{h}(x_j \oplus R_r^*)$

4.8 $G = h(s_j \parallel ID_k \parallel T_2 \parallel R_t^* \parallel h(s_j \oplus R_t^*))$

4.9 $\xrightarrow{E,F,G}$

4.10 $After\ successful\ authentication\ updates$

$x_{j-1} \leftarrow x_j;\ x_j \leftarrow h(x_j \oplus R_r)$

$s_{j-1} \leftarrow s_j;\ s_j \leftarrow s_j \oplus R_t$

$ID_k^{old} \leftarrow ID_k$

$ID_k^{new} \leftarrow ID_k \oplus s_j$

**Reader:**

5.1 $Compute$

$E^* = h(x_j \parallel RID_k \parallel T_1 \parallel R_r \parallel h(x_j \oplus R_r))$

5.2 $Check\ E^* \stackrel{?}{=} E$

5.3 Updating $x_j \leftarrow h(x_j \oplus R_r)$

5.4 $Data = F \oplus \mathrm{h}(x_j \oplus R_r)$

5.5 $\xrightarrow{G}$

**Tag:**

6.1 $Compute$

$G^* = h(s_j \parallel ID_k \parallel T_2 \parallel R_t \parallel h(s_j \oplus R_t))$

6.2 $Verify\ G^* \stackrel{?}{=} G$

6.3 $After\ successful\ authentication$

$s_j \leftarrow s_j \oplus R_t$

$ID_k \leftarrow ID_k \oplus s_j$

Fig. 3 Improved version of SRTA protocol.

through the protocol causes another vulnerability, i.e., DoS attack. In other words, after running four steps of the protocol successfully, the attacker intercepts the protocol and leaves it unfinished. So the back-end server updates $ID_k$ with $ID_k \oplus R_t$, while the value of $ID_k$ in the tag is not updated. Now in the next run of the protocol, the tag will send $ID_k$ to the reader but the back-end server will not admit it as a legitimate one. So, we store two values for $ID_k$ in the back-end server as a new and old ones. Moreover, we update $ID_k$ at the end of the protocol as follows:

$$ID_k \leftarrow ID_k \oplus s_j \qquad (16)$$

and stores two last value of $ID_k$ in the back-end server side. As we mentioned above, restriction of complexity in the tag side is an important issue, so by omitting one hash function in tag, we change the updated value of $ID$ as eq. 16. The improved protocol is depicted in Fig. 3.

## 6. Analyzes of our proposed protocol

**Eavesdropping and Tracing Resistance**

Our proposed protocol is resistant to eavesdropping and tracing attacks. An adversary is not able to trace the target tag $T_0$, because of updating $ID_k$ as $ID_k \oplus s_j$, in addition $s_j$ is updated at the end of protocol with $R_t$ which is generated randomly and is not known to the attacker $\mathcal{A}$. So the barrier against tracing is raised through the use of random numbers and anonymity.

**Desynchronization Attack Resistance**

In desynchronization attack, the adversary forces the tag and the reader to update their secret values to different ones. So, they will not authenticate each other in further transactions. In an RFID authentication protocol, the adversary can perform this attack via various approaches including blocking exchanged messages between the tag and the back-end server and impersonating the tag and the reader [29, 30]. In our protocol, an adversary is not able to forge the tag and the reader to update their secret values, because of storing two values of $ID_k$ in the back-end server, which prevent desynchronization between the tag and the back-end server.

Table 1. Security level comparisons among the discussed protocol

| Feature → / Protocols ↓ | $F_1$ | $F_2$ | $F_3$ | $F_4$ | $F_5$ |
|---|---|---|---|---|---|
| Cho et al. [21] | NO | YES | NO | NO | NO |
| Srivastava et al. [6] | NO | YES | NO | NO | YES |
| Li et al. [7] | YES | NO | NO | YES | NO |
| Our protocol | YES | YES | YES | YES | YES |

$F_1$: Provision of mutual authentication
$F_2$: Provision of synchronized secret
$F_3$: Protection of data privacy
$F_4$: Prevention of reader stolen/lost attack
$F_5$: Prevention of impersonation attack

Table 2. Performance features of various protocols

| Feature → / Protocols ↓ | complexity of tag computation | complexity of reader computation | Communication rounds |
|---|---|---|---|
| Srivastava et al. [6] | 5H+RNG | RNG | 5 |
| Li et al. [7] | 3H+RNG | RNG | 5 |
| Our protocol | 3H+RNG | RNG | 5 |

H hash function, RNG random number generator

**Tag/Reader impersonation Attack Resistance**

Tag (Reader) impersonation attack is a forgery attack, in which an RFID system accepts a spoofed tag (reader) as a legitimate tag (reader). In our improved protocol, because of the new exposure of $B$ and $D$, an adversary $\mathcal{A}$ is not able to build the messages $B$ and $D$ from $A$ and $C$. Furthermore, because of updating the secret values and generation of new random variables in each session, the eavesdropped messages from the last session are not acceptable in the new session.

### 6.1 Performance analysis of our proposed protocol

In Table 1, our improved protocol is compared with some similar protocols. As it can be seen, the proposed protocol provides security against the mentioned attacks including traceability, impersonation, mutual authentication and DoS. In addition, in Table 2, the efficiency of the proposed protocols is compared with the analyzed protocols, by comparing its computational cost. Moreover, qualitative values of our proposed protocol is evaluated over discussed pervious protocols.

## 7. Conclusion

RFID Technology is rapidly developing and its applications are spreading in different fields, but providing their security and privacy is the goal of researchers in recent years. In this paper, we analyzed a hash based RFID protocol in TMIS, proposed by Li et al.. They claimed that their protocol provides privacy requirements for RFID systems. However, this paper showed that Li et al.'s protocol is still vulnerable to traceability, tag impersonation and DoS attacks and to fix the aforementioned weaknesses, we have proposed an improvement, which fixes the weak features of their protocol for healthcare environments. Finally, the computational complexity and the performance of the proposed protocol is compared with discussed protocols.

## REFRENCE